\documentclass[12pt]{article}
\usepackage{latexsym,amssymb, amsthm, epsfig}
\usepackage{amsfonts,amsmath}
\newcommand{\Z}{{{\mathbb{Z}}}}
\newcommand{\R}{{{\mathbb{R}}}}

\newcommand{\wh}[1]{\widehat{#1}}

\newcommand{\ovl}[1]{\overline{#1}}
\newcommand{\reff}[1]{(\ref{#1})}

\newcommand{\rd}{{{\rm d}}}

\begin{document}

\title{Markov Process of Muscle Motors}

\author{ Yu. Kondratiev$^{1}$, E. Pechersky$^{2}$, S.
Pirogov$^{2}$\\
\footnotesize{$^{1}$ Bielefeld University, 33601 Bielefeld,
Germany}\\
\footnotesize{$^2$ IITP, 19, Bolshoj Karetny per., GSP-4, Moscow
127994, Russia }\\
\footnotesize{kondrat@mathematik.Uni-Bielefeld.DE, pech@iitp.ru,
pirogov@iitp.ru}}

\date{}
\maketitle

\vspace{15mm}

\begin{abstract}
We study a Markov random process describing a muscle molecular
motor behavior. Every motor is either bound up with a thin
filament or unbound. In the bound state the motor creates a force
proportional to its displacement from the neutral position. In
both states the motor spend an exponential time depending on the
state. The thin filament moves at its velocity proportional to
average of all displacements of all motors.

We assume that  the time which a motor stays at the bound state
does not depend on its displacement. Then one can find an exact
solution of a non-linear equation appearing in the limit of
infinite number of the motors.
\end{abstract}

\newpage

\section{Introduction}  Recent progress in the molecular motors
study brought  a wave of new models and methods in  theoretical
considerations. The tools involved into considerations are spread
from biochemistry and biophysics to mathematics and probability
theory. As examples, we would like to mention  works
\cite{J1,J2,J3,R, MW}. This  list contains only some papers from
different areas.

We concentrate here on mathematical aspects of the problem and,
more precisely, we describe  a probabilistic model  of the muscle
motor which leads to a non-linear Markov process. The latter
notion was introduced by E.P.Mc-Kean \cite{Mc}. Comparing with the
usual case, in a non-linear Markov process  transition
probabilities are related to a non-linear equation. With respect
to the probability theory, non-linear Markov processes play  main
role in the modelling. Because of non-linearity, the model appears
to have many effects atypical for usual Markov processes that
gives possibilities either to explain experimentally observed
properties of the motors or to predict new ones. Following Howard'
book \cite{How} we shall distinguish two classes of the molecular
motors: processive motors and non-processive ones. The distinction
between these classes leads to different types of the models,
however  differences are relative. Perhaps one can construct a
general model yielding all features of both sorts of motors. We
consider here a model of non-processive motors which concerns
motors involved in the muscle activity.

\section{Model} \subsection{Informal description} Main
components of a molecular motor complex are  protein molecules
called the motors themselves  which perform a motion along  a
molecule called filament (or microtubule) and playing a role of
railway for the motors. At last, there exists a cargo or back bond
which play passive role of  a relocatable ingredient. As it can be
extracted from the biochemical and biophysical literature, there
are several kinds of the motor protein molecules involved in
slightly different types of movements.

In this work we study the motors producing muscle motions. Those
motors are fixed by one of their end at a long protein molecule
called the thick filament. The thick filament is fastened in a
cell. Thus the motors do not move with respect to the cell they
are located in. The second end of the motors may be attached  or
not attached to a thin filament, another long molecule in the cell
disposed parallel to the thick filament. If both ends of a motor
are attached then the motor can cause a tension acting on the thin
filament if the end positions are not opposite each other. This
tension is the cause of the thin filament motion. The binding to
and unbinding out   the thin filament of any motor are random.
When binding a motor chooses a point on the thin filament to
attach randomly. The tension the motor creates is defined by a
displacement value between two ends of the motor. Therefore all
bounded motors create different tensions. It is common to take the
velocity of the thin filament moving proportional to the average
of all bounded filament tensions.

\subsection{Formal description} The formalization of the muscle motor
construction informally described above can be done in the
following way. We introduce a random process of interacted
particles. Every particle represent a motor. Therefore we locate
the particles at points of $\Z$.

Let us start with a random process $\zeta_{k}(t)$ describing the
binding and unbinding process the particle located at $k\in\Z$.
The state space is the two-point set $D=\{0,1\}$, where 0 means
the unbinding and 1 means the binding particle state. Then the
infinitesimal operator of $\zeta_{k}(t)$ is the following
$2\times 2$ matrix
\begin{equation*}
L_{k}^{D}=\begin{pmatrix} -c_{b}&c_{b}\\
c_{u}&-c_{u}
\end{pmatrix},
\end{equation*}
where $\Pr\big(\zeta_{k}(t)=1\big/
\zeta_{k}(0)=0\big)=c_{b}t+o(t)$ and $\Pr\big(\zeta_{k}(t)=0\big/
\zeta_{k}(0)=1\big)=c_{u}t+o(t)$.

Then the probability $P_{k}(t)$ of $k$h motor to be unbound at the
time $t$ satisfies the equation
\begin{equation}\label{equf}
\frac{\rd P_{k}(t)}{\rd t}=-c_{b}P_{k}(t)+c_{u}(1-P_{k}(t)),
\end{equation}
which gives in the steady state
\begin{equation}\label{steady}
P_{k}=\frac{c_{u}}{c_{b}+c_{u}}.
\end{equation}

The state space $\mathbb{X}$ of $k$h motor consists of pairs
$(z,\varepsilon)\mbox{ where } z\in\mathbb{R}\mbox{ if
}\varepsilon=1$ and $z=0 \mbox{ if }\varepsilon=0$. As before the
parameter $\varepsilon$ indicates the bounded (if $\varepsilon=1$)
and unbounded (if $\varepsilon=0$) positions of the particle. We
define a random process $\xi_{k}(t)$ describing binding and
unbinding actions and the displacement when binding, including the
deterministic moving of a single particle. Let $b(x)$ be a
distribution density such that $\int xb(x)\rd x>0$. The
infinitesimal operator of the process defining the behaviour of
$k$th particle is
\begin{align}\label{genmov}
L_{k}f(x,\varepsilon)=c_{b}\left[\int b(z)f(z,1-\varepsilon)\rd
z-f(x,\varepsilon)\right](1-\varepsilon)-\\
\varkappa x\frac{\rd}{\rd x}f(x,\varepsilon)\varepsilon+
c_{u}\left[f(k,1-\varepsilon)-
f(0,\varepsilon)\right]\varepsilon\nonumber,
\end{align}
Now a constant $c_{b}>0$ is the rate of the particle to jump to a
point of $\R\times\{1\}$ from the state $(0,0)$, that is
$\Pr\big(\zeta_{k}(t)\in\R\times\{1\}\big/
\zeta_{k}=(0,0)\big)=c_{b}t+o(t)$. A constant $c_{u}>0$ is the
rate of the particle to jump to $(0,0)$ from  any point in
$\R\times\{1\}$. That means that the rate of unbinding does not
depend on the point $x\in\R$ where the particle was attached at
the unbinding moment.   The function $b(x)$ is the probability
density of the  particle to bind at the point $x$ if the particle
was at $(0,0)$. Here $x$ is the displacement of the particle with
respect to its neutral position. Constant $\varkappa$ is positive.
It is the tension of the single motor molecule. It can be seen
from \reff{genmov} that when the particle is on $\R$, that is
$\varepsilon=1$, then it is moving to the point $0$ with the
velocity proportional to $x$.

For the density $p_{k}(x,t)$ of the particle to be on $\R$ we
obtain the following differential equation (see \cite{GiSko})
\begin{equation}\label{equ1}
\frac{\partial p_{k}(x,t)}{\partial
t}=c_{b}b(x)P_{k}(t)+\varkappa\frac{\partial}{\partial x}
\big[xp_{k}(x,t)\big]-c_{u}p_{k}(x,t),
\end{equation}
where $P_{k}(t)=1-\int p_{k}(x,t)\rd x$. The probability
$P_{k}(t)$ to be unbound  satisfies \reff{equf}.

Next we introduce the interaction between the particles. We
cannot  express the interaction in a Hamiltonian form.  Instead
we introduce a deterministic dynamic of all bound particles such
that particles dynamic is highly correlated each to other.
Moreover all particles are moving with the same velocity. To be
more precise consider all particles in the interval
$[-N,N]\subset\Z$ and the configuration space
$\Omega_{N}=\mathbb{X}^{[-N,N]}$ of all particles in $[-N,N]$.
The space $\Omega_{N}$ is a disjoint union of the sets
$\Omega_{\varepsilon_{-N},...\varepsilon_{N}}
=\prod_{i}\R^{\varepsilon_{i}}$. We can consider any probability
distribution on $\Omega_{N}$ as a collection of measures on
spaces $\Omega_{\varepsilon_{-N},...\varepsilon_{N}}$.

The generator of the process involving all particles from
$[-N,N]$ is
\begin{align}
&L_{[-N,N]}f\big((x_{k},\varepsilon_{k}),k=-N,...,N\big)=\nonumber\\
&c_{b}\sum_{k=-N}^{N}\left[\int
b(z)f\big(...,(x_{k-1},\varepsilon_{k-1}), (z,1-\varepsilon_{k}),
(x_{k+1},\varepsilon_{k+1}),...\big)\rd
z-\right.\nonumber\\
&\left.f\big((x_{k},\varepsilon_{k}),k=-N,...,N\big)
\vphantom{\int}\right](1-\varepsilon_{k})-
\label{genmov}\\
&+c_{u}\sum_{k=-N}^{N}
\left[\vphantom{\int}f\big(...,(x_{k-1},\varepsilon_{k-1}),(k,1-\varepsilon_{k})
(x_{k+1},\varepsilon_{k+1}),... \big)\right.\nonumber\\
&\left.-f\big((x_{k},\varepsilon_{k}),k=-N,...,N\big)\vphantom{\int}
\right]\varepsilon_{k}-\nonumber\\
& v_{N} \sum_{k=-N}^{N}\frac{\partial} {\partial
x_{k}}f\big((x_{k},\varepsilon_{k}),k=-N,...,N\big)\varepsilon_{k},\nonumber
\end{align}
where
$v_{N}=\varkappa\frac{1}{2N+1}\sum_{k=-N}^{N}x_{k}\varepsilon_{k}
-{F}$. The term  $F$ in above formula means the velocity which an
external force adds to the common velocity
$\wh{v}_{N}=\frac{1}{2N+1}\sum_{k=-N}^{N}x_{k}\varepsilon_{k}$ of
all particles.

Further, we use the following notations. Let
$(\varepsilon_{-N},...,\varepsilon_{N})$ be fixed. Then
$M^{0}=\{i:\:-N\leq i\leq N, \varepsilon_{i}=0\}$ and
$M^{1}=(M^{0})^{c}=\{i:\:-N\leq i\leq N, \varepsilon_{i}=1\}$.

From now we shall denote the vector
$((x_{k},\varepsilon_{k}),k=-N,...,N)$ by $X$.

For a configuration $X=((x_{k},\varepsilon_{k}),k=-N,...,N)$ and
$i\in M^{1}$ define the configuration $u_{i}X$ for which the pair
$(x_{i},1)$ is substituted by $(0,0)$; for $i\in M^{0}$ and $x\in
\R$ define $b_{i}^{x}X$ as the configuration for which
$(0,\varepsilon_{i}=0)$ is substituted by $(x,1)$.

Let $p^{N}(X,t)$ be the measure density  of all bound particles to
be at the given points  at the time moment $t$.  Then
$p^{N}(X,t)$ satisfies the equation
\begin{align}
&\frac{\partial p^{N}(X,t)}{\partial
t}+\sum_{i=-N}^{N}\frac{\partial}{\partial
x_{i}}\Big[\varepsilon_{i}v_{N} p^{N}(X,t)\Big]
=\nonumber\\
&c_{b}\sum_{i=-N}^{N}\varepsilon_{i} b(x_{i})p^{N}(u_{i}X,t)+
c_{u}\sum_{i=-N}^{N}(1-\varepsilon_{i})\int
p^{N}(b_{i}^{x}X,t)\rd x
-\label{equ}\\
&\left[c_{u}\sum_{i=-N}^{N}\varepsilon_{i}+
c_{b}\sum_{i=-N}^{N}(1-\varepsilon_{i})\right]p^{N}(X,t).\nonumber
\end{align}

Let $p_{k}^{N}(x,t)$ be the probability density  of the $k$-th
particle to be at the point $x$ at the time moment $t$.

We  are interested in the behavior of the motor system for the
large number of motors which formally corresponds to the limit
$N\to\infty$. In this limit we can substitute $v_{N}$ by its
expectation value
\begin{equation}\label{velosity}
v=\frac{\varkappa}{2N+1}\sum_{i=-N}^{N}\int xp_{k}^{N}(x,t)\rd
x-{F}
\end{equation}

Plugging in $v$ into \reff{equ} we get the nonlinear equation
corresponding to non-linear Markov process describing interacting
motors.

Let us denote by
\begin{equation*}
\nu_{X}(\rd x)=\frac{1}{2N+1}\sum_{i=-N}^{N}\varepsilon_{i}
\delta_{x_{i}}(\rd x)
\end{equation*}
the random measure describing the motor distribution on the thin
filament. The expectation of this random measure is called the
first correlation measure and its density is called the first
correlation function, $n(x,t)$. It is evident that
\begin{equation}\label{corr}
n(x,t)=\frac{1}{2N+1}\sum_{k=-N}^{N}p_{k}^{N}(x,t).
\end{equation}
For the non-linear Markov process defined above the first
correlation function $n(x,t)$ satisfies the equation
\begin{equation}\label{how}
\frac{\partial n(x,t)}{\partial t}+v\frac{\partial
n(x,t)}{\partial x} =c_{b}b(x)(1-{N}(t))-c_{u}n(x,t),
\end{equation}
where $N(t)=\int n(x,t)\rd x$ and $v$ as above is
\begin{equation}\label{vel}
v=-\varkappa\int xn(x,t)\rd x+{F}
\end{equation}

It follows from \reff{how} that $N(t)$ and $v(t)$ satisfy the
equations
\begin{align}\label{inegr}
&\dot N=c_{b}(1-N)-c_{u}N\\
&\dot v= -\varkappa(vN+c_{b}(1-N)m_{1})+c_{u}(F-v),\nonumber
\end{align}
where $m_{1}=\int xb(x)\rd x$. Evidently for $t\to\infty$ $N$ and
$v$ tend to their limit values

\begin{align}\label{PP}
&\ovl{N}=\frac{c_{b}}{c_{b}+c_{u}},\\
&\ovl{v}=-\frac{\varkappa\ovl{N}m_{1}-F}{1+\varkappa\frac{\ovl{N}}{c_{u}}}=
-\frac{\varkappa
c_{b}m_{1}-F(c_{b}+c_{u})}{c_{b}+c_{u}+\varkappa\frac{c_{b}}{c_{u}}}
\nonumber
\end{align}

\vspace{1cm}

Let us study the dependence of $\ovl{v}$ on $c_{u}$ and $c_{b}$.
It is clear that $\ovl{v}\to 0$ as $c_{u}\to 0$, and $\ovl{v}\to
F$ as $c_{u}\to \infty$. If $\varkappa m_{1}-F>0$ then $\frac{\rd
\ovl{v}}{\rd c_{u}}<0$ at $c_{u}=0$ and hence there exists a value
$c_{u}^{o}$ where $\ovl{v}$ is negative and  maximal in absolute
value $|\ovl{v}|=v^{o}$. It is achieved at
\begin{equation}\label{optcu}
c_{u}^{o}=\sqrt{\frac{F^{2}}{m_{1}^{2}}+c_{b}\frac{\varkappa
m_{1}-F}{m_{1}}} -\frac{F}{m_{1}}
\end{equation}
If there is no the external force $F=0$ then
$c_{u}^{o}=\sqrt{c_{b}\varkappa}$ and
\begin{equation}\label{maxi}
v^{o}=\frac{\sqrt{c_{b}}m_{1}}{2\sqrt{\varkappa}+\sqrt{c_{b}}}.
\end{equation}
In the case $F>0$ the velocity becomes positive for large the
unbound intensity $c_{u}>\frac{c_{b}(\varkappa m_{1}-F)}{F}$.

Recall that all above values was obtained under the condition of
small external force $F<\varkappa m_{1}$. If $F>\varkappa m_{1}$
then the velocity is positive at any $c_{u}$.

\vspace{1cm}

Let us consider more general case when the returning force
depends on $x$ non-linearly, so instead of \reff{vel} we have
\begin{equation}\label{nel}
v=\int\varphi(x)n(x,t)\rd x+F,
\end{equation}
where $\varphi(x)$ is some non-linear function.

\noindent Now we  substitute the expression \reff{nel} in the
equation \reff{how}. The expression for the time derivative of
$v$ is
\begin{equation}\label{ex}
\dot v=\int\varphi\dot n\rd x=-v\int\varphi\frac{\partial
n}{\partial x}\rd x+c_{b}(1-N(t))\int\varphi b(x)\rd x-c_{u}(v-F).
\end{equation}
Integrating by parts we obtain from the first term
$v\int\varphi^{'}(x)n(x,t)\rd x$. Inserting the new variable
$w=\int\varphi^{'}(x)n(x,t)\rd x$ we have $\dot w=
\int\varphi^{'}(x)\dot n(x,t)\rd x$ and using \reff{how} again we
get an expression for $\dot w$ containing the integral
$\int\varphi^{''}(x)n(x,t)\rd x$.

If $\varphi(x)$ is a polynomial then repeating this procedure we
finally obtain a finite system of ordinary differential equations.

The similar substitution is valid if $\varphi(x)$ is any
trigonometric polynomial or more generally any linear combination
of quasi-polynomials (i.e. usual polynomials multiplied by
sinusoidal and exponential functions \cite{Pont}). Consider the
simplest case $\varphi(x)=-\varkappa\sin(\alpha x)$. Denoting
$w=\int\cos(\alpha x)n(x,t)\rd x$ and $m_{c}=\int\cos(\alpha
x)b(x)\rd x$, $m_{s}=\int\sin(\alpha x)b(x)\rd x$ we have
\begin{align}
\dot N=&c_{b}(1-N)-c_{u}N\nonumber \\
\dot v=&-\varkappa\alpha v w - \varkappa
c_{b}(1-N)m_{s}+c_{u}(F-v)\label{sinegr}\\
\dot w=&\frac{\alpha
v(v-F)}{\varkappa}+c_{b}(1-N)m_{c}-c_{u}w.\nonumber
\end{align}
For any stationary point of this system the value $N$ is given by
\reff{PP} as before, and $v$ and $w$ can be found from $\dot
v=\dot w=0$. It is interesting to find the number of the
stationary points of \reff{sinegr}. Also it is interesting to
study the limit cycles for this system if they exist.

For the case $\varphi(x)=-\varkappa\sinh(\alpha x)$ the equation
for $N, v, w$ are similar to \reff{sinegr}  with the only
difference in the sign before the term $\frac{\alpha
v(v-F)}{\varkappa}$.

\section*{Acknowledgements}

The authors thank P. Reimann for valuable discussions on molecular
motors. Financial support of the SFB-701, Bielefeld University, is
gratefully acknowledged. E.P. and S.P. were partially supported by
the CRDF Grant RUM1-2693--MO-05.

\vspace{10cm}

\end{document}